\documentclass[twoside]{article}
\usepackage[dvips]{graphicx}
\input{mplal.sty}
\newcommand{\be}{\begin{equation}}
\newcommand{\ee}{\end{equation}}
\newcommand{\een}{\end{enumerate}}
\newcommand{\ben}{\begin{enumerate}}
\newcommand{\ber}{\begin{eqnarray}}
\newcommand{\eer}{\end{eqnarray}}
\newcommand{\pul}{\frac{1}{2}}
\newcommand{\nn}{\nonumber}

\newcommand{\bk}{\mbox{\boldmath$k$}}

\newcommand{\LA}{{\cal L}}
\newcommand{\M}{{\cal M}}
\newcommand{\ep}{\varepsilon}

\begin{document}
\runninghead{Tom\'{a}\v{s} Bahn\'{\i}k \& Ji\v{r}\'{\i} Ho\v{r}ej\v{s}\'{\i}}
{Deviations from low-energy theorem $\ldots$}
\normalsize\textlineskip
\thispagestyle{empty}
\setcounter{page}{1}


\vspace*{0.88truein}

\fpage{1}
\centerline{\bf DEVIATIONS FROM LOW-ENERGY THEOREM FOR $V_{L}
V_{L}$}
\vspace*{0.035truein}
\centerline{\bf SCATTERING DUE TO PSEUDO-GOLDSTONE BOSONS}
\vspace*{0.37truein}
\centerline{\footnotesize
Tom\'{a}\v{s} Bahn\'{\i}k$^{*}$\footnotetext{$^{*}$ e-mail: tomas.bahnik@vslib.cz}}
\vspace*{0.015truein}
\centerline{\footnotesize
\it Department of Physics, Technical University Liberec,}
\baselineskip=10pt
\centerline{\footnotesize\it H\'{a}lkova 6, 461 17 Liberec, Czech
Republic}
\vspace*{10pt}
\centerline{\footnotesize
Ji\v{r}\'{\i}
~Ho\v{r}ej\v{s}\'{\i}$^{\dagger}$\footnotetext{$^{\dagger}$ e-mail:
jiri.horejsi@mff.cuni.cz}}
\vspace*{0.015truein}
\centerline{\footnotesize\it
Nuclear Centre, Faculty of Mathematics and Physics,
Charles University,}
\baselineskip=10pt
\centerline{\footnotesize\it V Hole\v{s}ovi\v{c}k\'{a}ch 2,
180 00 Prague 8, Czech Republic}
\vspace*{0.225truein}

\vspace*{0.21truein}
\abstracts{
Possible deviations from a low-energy theorem for the scattering of
strongly interacting longitudinally polarized $W$ and $Z$ bosons are
discussed within a particular scheme of electroweak symmetry breaking.
The scheme (suggested earlier by other authors in a slightly different
context) is based on spontaneous breakdown of a $SU(4)$ symmetry to
custodial $SU(2)$ subgroup. The physical spectrum of such a model
contains a set of relatively light pseudo-Goldstone bosons whose
interactions with vector bosons modify the low-energy theorem proven for
a ``minimal'' symmetry-breaking sector. The Goldstone-boson manifold
$SU(4) / SU(2)$ is not a symmetric space. In this context it is observed
that, on the other hand, there is a large class of models of electroweak
symmetry breaking, involving groups $G$ and $H$ such that the $G/H$ is a
symmetric space and the corresponding rich multiplets of
pseudo-Goldstone bosons do not influence the canonical low-energy
theorem. For the scheme considered here, the relevant interactions are
described in terms of an effective chiral Lagrangian and tree-level
contributions of the pseudo-Goldstone boson exchanges to the vector
boson scattering are computed explicitly. A comparison with the Standard
Model is made.}{}{}

\vspace*{1pt}\textlineskip      
\section{Introduction}
\vspace*{-0.5pt}
\noindent
The nature of electroweak symmetry breaking (EWSB), i.e. the
mechanism responsible for generating the $W$ and $Z$  boson masses,
constitutes perhaps the most important open issue of the
present-day particle physics.\cite{Peskin97} One of the few
(indirect) experimental clues in this respect seems to be
provided by the value of the famous parameter
$\rho = m_{W}^{2} / \left( m_{Z}^{2} \cos^{2} \theta_{W} \right)$,
which is known to be close
to unity. Such a value is understood quite naturally within
spontaneously broken $SU(2) \times U(1)$ gauge theories if an unbroken
``custodial'' $SU(2)$ symmetry is present, giving automatically $\rho =
1$ in the lowest order.\cite{Sikivie} Thus, the Goldstone realization
of an appropriate internal symmetry of a (largely unknown)
``symmetry-breaking sector'' and the ensuing Higgs mechanism (in a
most general sense) seem to be highly plausible generic features
of a realistic theory of electroweak interactions. Needless to
say, the successful minimal standard model (SM) is built
precisely along these lines, under an additional (technical)
constraint of perturbative renormalizability and with the symmetry of
the Higgs system being $O(4) \simeq SU(2) \times SU(2)$. On the other
hand, given only the fairly weak phenomenological restrictions
mentioned above, one may clearly envisage more general theoretical
schemes for EWSB.

It is well known that the conceivable EWSB scenarios are
basically of two distinct types. First, one may consider a
(rather general) multiplet of elementary Higgs scalar fields and
a corresponding ``potential'', whose specific form triggers the
spontaneous symmetry breakdown; one thus follows essentially the
SM paradigm and perturbative renormalizability is maintained.
Such a perturbative approach makes sense if the relevant
couplings are sufficiently weak and this in turn means that a
physical Higgs boson becomes relatively light, having mass far
below 1 TeV or so.\cite{HHG,Espinosa} Second, there is a
``strong-coupling scenario'' (see e.g. Refs.~5,~6
for a review), in which EWSB is assumed to have a
similar origin as e.g. the spontaneous breakdown of chiral
symmetry within QCD. The corresponding new strong interaction
responsible for EWSB may then be labelled generically e.g. as
``technicolour''\cite{Farhi} (though we will actually not invoke any
particular dynamical model here). In a minimal variant of such a
scheme there are just three (dynamically generated) Goldstone
bosons that are used up in the Higgs mechanism, giving rise to
the longitudinal polarizations (i.e. zero helicities) of massive
vector bosons $W$ and $Z$; in the physical spectrum there is then no
natural counterpart of the SM Higgs boson. Of course, one thus
also abandons the concept of conventional renormalizability.
There is a typical mass scale $\Lambda_{SB}$ of the symmetry-breaking (SB)
sector and for sufficiently low energies
($E < \Lambda_{SB}$) an
appropriate description of the dynamics is provided in terms of
an effective ``chiral'' Lagrangian (cf. e.g. Refs.~8,~9
for review), in analogy with phenomenological pion
Lagrangians in low-energy QCD.\cite{Weinberg,GL} In this context,
it should be noted that the current SM precision tests and
analyses of electroweak radiative corrections suggest that the
perturbative scenario with a light Higgs boson is favoured over
the simplest technicolour models.\cite{Altarelli} Nevertheless, from
a more general perspective, a strongly interacting SB sector
within the $SU(2) \times U(1)$ electroweak gauge theory still represents
a viable alternative -- a clear-cut answer concerning the EWSB
issue can only come from results of the future experiments at LHC
or elsewhere.\cite{ChanowitzPHP,Dominici}

For an electroweak theory with strongly interacting SB sector a
remarkable general result is valid, namely a low-energy theorem (LET)
concerning the scattering of longitudinally polarized vector
bosons.\cite{ChanowitzLET} This theorem specifies a universal leading
behaviour of the $V_{L}V_{L}$ scattering amplitudes (the $V$ is a common
symbol for $W$ or $Z$ ) for energies such that $m_{W}^{2} \ll s \ll
\Lambda_{SB}^{2}$. Roughly speaking, one observes linear growth with $s
= E_{c.m.}^{2}$, characteristic for the $SU(2) \times U(1)$ gauge
interactions alone (i.e. without the additional damping provided by a
light Higgs boson within SM). Note also that the LET formulated by
Chanowitz et al.\cite{ChanowitzLET} is analogous to the low-energy
results known for pion-pion scattering\cite{Weinberg,GL,Donoghue}; such
a formal similarity is due to the famous equivalence theorem\cite{ET}
for Goldstone bosons and longitudinal vector bosons. The LET has been
proved in Ref.~14 under the assumption that all would-be Goldstone
bosons are ``eaten'' by $W$ and $Z$ resp. as a result of the Higgs
mechanism, i.e. that the $W$ and $Z$ are the only light particles in the
physical spectrum (``light'' means a mass comparable with $m_{W}$ or
less). In other words, the assumed symmetry pattern corresponds to
$SU(2) \times SU(2)$ broken down to a (custodial) $SU(2)$, or simply
$SU(2) \times U(1)$ down to the electromagnetic $U(1)$ (in the latter
case the relation $\rho = 1$ is not automatic and must eventually be
imposed by hand).

Of course, one is allowed to consider more general SB schemes,
without obviously violating the existing phenomenology. A
question then immediately arises as to whether (and how) the LET
may get modified. A general pattern of symmetry breakdown can be
characterized by specifying the symmetry group $G$ of the SB sector
and its unbroken subgroup $H$ (in particular, the $H$ may be
conveniently chosen so as to contain a custodial $SU(2)$).
Goldstone bosons associated with spontaneous
breaking of the symmetry  $G$ down to $H$
correspond to coordinates in the manifold
(quotient space) $G/H$ and one can write a corresponding effective
Lagrangian for the SB sector (involving a nonlinear realization
of the symmetry $G$) in a model-independent way, employing the
general CCWZ construction.\cite{CCWZ} When a subgroup $G_{w} = SU(2)
\times U(1)$ of the $G$ is gauged, three Goldstone bosons disappear, but
some other may survive (if $G/H$ is large enough) in the physical
spectrum and acquire masses through gauge interactions. In
general, such pseudo-Goldstone bosons (PGB) may be relatively
light, i.e., have masses on the electroweak scale and thus they
are expected to modify the LET for $V_{L} V_{L}$ scattering
through the additional exchanges in the corresponding tree-level
Feynman diagrams.\cite{ChanowitzLET}

The purpose of the present paper is to examine such deviations from the
LET quantitatively, in a pertinent model and using the corresponding
effective Lagrangian. To this end, one might consider first e.g. the
schemes discussed earlier by various authors in the context of explicit
technicolour models.\cite{Peskin,Chadha,Preskill} These examples
include the SB patterns with $G/H = SU(n) \times SU(n) \times U(1) /
SU(n) \times U(1)$, $SU(2n)/O(2n)$ and $SU(2n)/Sp(2n)$. Note that in all
these cases, the quotient space $G/H$ is a symmetric space (i.e. there
is a natural ``parity'' operation\cite{CCWZ} distinguishing the broken
and unbroken generators of the $G$). Looking into the interactions of
the corresponding physical PGB (see in particular Ref.~18 ), one finds
that there are no modifications of the LET in question -- interactions
of the type $\pi V V$ (where $\pi$ stands for a PGB) become trivial
within all these schemes.

It turns out that a minimal scheme suitable for our purpose is
provided by the model suggested some years ago by Chivukula and
Georgi.\cite{ChiG} The corresponding SB pattern is $G/H = SU(4) /
SU(2)$ and does lead to non-trivial $\pi V V$ interaction vertices; the
study of deviations from the LET for $V_{L} V_{L}$ scattering
within such a scheme is the main subject of our investigation. It
should be stressed that throughout our discussion we stay within
the general framework of the CCWZ effective Lagrangian and do not
attempt to find any particular dynamical model (technicolour or
whatever) producing the SB pattern in question. As a side remark,
let us also note that another interesting feature of the
considered scheme is that it contains doubly charged PGB; in this
respect, it has a similar signature as some particular
weak-coupling models involving Higgs triplets (in addition to a
standard doublet), discussed independently in the literature
by several authors.\cite{Machacek,Vega}

The present paper is organized as follows. In the next section
the statement of the LET for $V_{L} V_{L}$ scattering is briefly
summarized. In Sect.3 the $SU(4) / SU(2)$ EWSB scheme is described in
some detail and Sect.4 is devoted to the calculation of the
envisaged deviations from the LET due to the PGB exchange.
Main results of the calculation are discussed briefly in Sect.5,
while some further remarks of a more general character are left
to Sect.6. Some uncomfortably long formulae employed in our
calculation are relegated to an Appendix.

\section{Low-Energy Theorem for $V_{L} V_{L}$ Scattering\label{sec:let}}
\noindent
As noted in the Introduction, the LET due to Chanowitz et
al.\cite{ChanowitzLET} specifies the leading energy-dependence of the
$V_{L} V_{L}$ scattering amplitudes within models involving a
``minimal'' strongly interacting SB sector, i.e. such that the only
Goldstone bosons are those ``eaten'' by the $W^{\pm}$ and $Z$ via the
Higgs mechanism. Here we will recapitulate briefly only the main
results; for more details, the reader is referred to the original
literature.

The theorem should hold for the energy domain
$m_{W}^{2} \ll s \ll \Lambda_{SB}^{2}$, with
\begin{equation}
\label{nova}
\Lambda_{SB} = \mbox{min} \left\{ 4\pi v, M_{SB} \right\} \sim 1
\mbox{TeV},
\end{equation}
where $v$ is the well-known electroweak scale,
$v = \left( G_{F} \sqrt{2} \right)^{-1/2} \doteq 246 \; \mbox{GeV}$
($G_{F}$ being
the Fermi constant) and $M_{SB}$ denotes a mass scale typical
for the SB sector (e.g. the mass of the lightest resonance due to
the strong EWSB force -- a ``technirho'' or so). The essential
statement of the LET (up to the order $O \left( g^{2} \right)$
in electroweak gauge
coupling) is summarized in Table 1. Upon neglecting small
constant terms, the LET expressions coincide with the high-energy
($E \gg m_{W}$) limit of the gauge part of the complete
amplitude -- for a detailed calculation see e.g. Ref.~23
(cf. also Ref.~24 ). The leading behaviour shown in Table~1
is universal, i.e. independent of any particular dynamical model
reproducing the considered SB pattern.

\begin{table}[ht]
\tcaption{Leading terms in the $V_{L} V_{L}$ scattering amplitudes
predicted by the LET, $\rho = m_{W}^{2} / \left( m_{Z}^{2}
\cos^{2} \theta_{W} \right)$\label{tab:let}}
\begin{center}
\begin{tabular}{|c|c|}\hline
\rule[-3mm]{0cm}{8mm} Process & LET prediction \\ \hline\hline
\rule[-3mm]{0cm}{8mm} $W^+_L W^-_L \to Z_L Z_L$ & $\frac{g^2 s}{4 \rho m_W^2}$
\\ \hline
\rule[-3mm]{0cm}{8mm} $W^+_L Z_L \to Z_L W^+_L$ &$\frac{g^2 u}{4 \rho m_W^2}$
\\ \hline
\rule[-3mm]{0cm}{8mm} $W^+_L W^+_L \to W^+_L W^+_L$
& $-\frac{g^2 s}{4 m_W^2}\left(4 - \frac{3}{\rho}\right )$
\\ \hline
\rule[-3mm]{0cm}{8mm} $W^+_L W^-_L \to W^+_L W^-_L$
&$-\frac{g^2 u}{4 m_W^2}\left(4 - \frac{3}{\rho}\right)$ \\ \hline
\rule[-3mm]{0cm}{8mm} $Z_L Z_L \to Z_L Z_L$
&0 \\ \hline
\end{tabular}
\end{center}
\end{table}

The analysis of Ref.~14 shows that under the
assumptions mentioned above all corrections to the scattering
amplitudes from strongly interacting EWSB sector decouple, except
for a renormalization of the $m_{W}$ and $\rho$. The LET has
been derived there by means of three different methods,
in particular via perturbative Feynman-diagram analysis, then
with the help of an appropriate effective Lagrangian and also by
using current-algebra techniques -- the latter approach stresses
an analogy with pion low-energy theorems.\cite{Weinberg,GL,Donoghue}

\newpage
\section{The $SU(4)/SU(2)$ Symmetry-Breaking Scheme\label{sec:su4su2}}
\noindent
We will employ here the SB scheme proposed some years ago by
Chivukula and Georgi\cite{ChiG}, in which the original symmetry
group $G$ of a strongly interacting SB sector is $SU(4)$ and its
unbroken subgroup $H = SU(2)$ (playing the role of the
custodial symmetry). No particular details of possible dynamics that
might produce such a SB pattern are assumed and the relevant
interactions are described in terms of a model-independent
effective Lagrangian constructed according to the known general
principles.\cite{CCWZ} The Lagrangian contains twelve Goldstone bosons;
three of them are absorbed by the Higgs mechanism and the remaining ones
become massive, with masses of order $\le 1\,\rm TeV$ depending on the
parameters of the Lagrangian. In particular, the scheme involves doubly
charged PGB that couple directly to two similarly charged $W$'s. These
make interesting contributions to $W^+ W^+$ and $W^+ W^-$ scattering
amplitudes\footnote{There is also a weak-coupling
scheme\cite{Machacek,Vega} with extended Higgs sector including
multiplets classified under custodial $SU(2)$ as a quintuplet
$H^{++,+,0,-,--}_5$, a triplet $H^{+,0,-}_3$ and two singlets $H^0_1$
and $H'_1$.}. In the remainder of this section we summarize the
essential properties of the relevant effective Lagrangian; some further
details can be found in the original paper Ref.~20 and also in Ref.~23.

Let us denote by ${\cal SU}(4)$ and ${\cal SU}(2)$ resp. the Lie algebras of
the initial global symmetry group $SU(4)$ and its unbroken  subgroup
$SU(2)$ resp. The generators of ${\cal SU}(4)$ are decomposed as $\lambda_{A}
= \{t_i, x_a \}$ where the $t_i$ form a basis in the unbroken ${\cal
SU}(2)$ and the $x_a$ are broken generators, i.e. correspond to a basis in the
orthogonal subspace. Generators are normalized in such a way that $
\mbox{Tr}\{\lambda_A\lambda_B^{\dagger}\} = \delta_{AB} $. According to general
arguments\cite{CCWZ} the number of Goldstone bosons (GB) is
equal to the number of broken generators, $n_{GB} = {\rm dim}\,{\cal
SU}(4) - {\rm dim}\,{\cal SU}(2) = 12$. The GB, or broken generators,
may in general be decomposed into irreducible multiplets with
respect to the unbroken subgroup.
In the considered case we have one GB quintuplet $\pi^{5}_{++,+,0,-,--}$, two
triplets $\pi^{3}_{+,0,-}$, $\pi^{3'}_{+,0,-}$ and one  singlet
$\pi^{1}$ under the ${\cal SU}(2)$. The corresponding effective
Lagrangian consists of three parts
\be {\cal L} =
{\cal L}_{IVB}  +  {\cal L}_{KE} + {\cal L}_{m}  \ .
\label{eq:genlag}
\ee
The first term represents the usual gauge boson interactions (as
well as their kinetic terms)
\be
\LA_{IVB}=-\frac{1}{4}F^i_{\mu\nu} F^{i\mu\nu} -\frac{1}{4} G_{\mu\nu}
G^{\mu\nu} \ .
\ee
The ${\cal L}_{KE}$ is the gauge invariant kinetic term of GB
fields
\ber
{\cal L}_{KE}& = & \pul f_5^2 J^{5\mu}\cdot J^{5*}_{\mu} +
\pul f_3^2 J^{3\mu} \cdot J^3_{\mu} \nn\\ &  & + \pul f_{3'}^2 J^{3'\mu}
\cdot J^{3'}_{\mu} + \alpha f_3 f_{3'} J^{3\mu}\cdot J^{3'}_{\mu} \nn\\
&  & + \pul f_1^2 J^{1\mu} J^1_{\mu} \ .
\label{lke}
\eer
The $J_{\mu}^j$
are defined by a decomposition of the quantity $e^{-i\Pi}(\partial_{\mu} +
i W_{\alpha\mu}w_{\alpha})e^{i\Pi}$ into broken and unbroken subspaces
\be
e^{-i\Pi}(\partial_{\mu} + i W_{\alpha\mu}w_{\alpha})e^{i\Pi} = i
\sum_{j=1}^M J_{\mu}^j\cdot x^j + T_{\mu}
\ee
where the superscript $j = 1,3,3',5$ labels irreducible multiplets and
the dot denotes a ``scalar'' product inside given
multiplet $j$
\be
J_{\mu}^j\cdot x^j = \sum_{a=1}^{n_j} J_{\mu a}^j x^j_a
\ee
The GB fields are contained in
\be \Pi = \sum_{j=1}^M \frac{\pi^j \cdot
x^j}{f_j}
\ee
where $f_j$ are parameters with dimension of a mass (they are
analogous to the pion decay constant in the low-energy chiral
Lagrangian of QCD).
The expression $\partial_{\mu} + i W_{\alpha\mu}w_{\alpha}$ is ordinary
gauge covariant derivative, i.e. $W_{\mu\alpha}$ are gauge boson
fields and the $w_{\alpha}$ denote generators of a $SU(2)\times U(1)$ electroweak
gauge group including gauge coupling constants $g$ and $g'$.

The last term parametrizes electroweak contributions to the GB scattering
of the order $g^2$ and/or $g'^2$. It contains mass terms and nonderivative
self-interactions of the GB. Using this part of the Lagrangian we can identify
fictitious GB, eaten in the Higgs mechanism, real GB, which remain
massless and the PGB with masses induced by explicit symmetry
breaking given by electroweak interaction\footnote{We adopt here
the terminology of Ref.~25.}. The determination of the
$\LA_m$ is known as the vacuum alignment
problem.\cite{Peskin,Chadha,Preskill} Within the
scheme under consideration we have just three massless GB (namely the
fictitious ones) and nine massive PGB.\cite{ChiG,bahnik}

The final form of the $\LA_m$ reads
\ber -\LA_m = V(\Pi)& = &(a + 2|d|) f^4 \,Tr\{w_\alpha(\Pi)x^5\}\cdot
Tr\{w_\alpha(\Pi)x^{5\dagger}\} \nonumber \\ &+ & (b + 2|d|)
f^4\,Tr\{w_\alpha(\Pi) x^{3'}\}\cdot Tr\{w_\alpha x^{3'}\} \nonumber\\
&+&(c + 2|d|) f^4\,Tr\{w_\alpha(\Pi) x^1\} Tr\{w_\alpha(\Pi)
x^1\}\nonumber\\ &+ & d\, f^4\, Tr\{w_\alpha(\Pi) \sigma_a \times 1\}
Tr\{w_\alpha(\Pi) 1 \times \tau_a\} \ \nn
\eer
where the $x^j$ denote multiplets of broken generators, $\sigma$ and $\tau$ are
Pauli matrices and $w_\alpha(\Pi) = e^{-i\Pi}w_{\alpha}e^{i\Pi}$.
Parameters $a,b,c$ are positive dimensionless numbers of the order
of unity and
$f$ has dimension of a mass. It is expected that different $f$'s,
i.e. the $f_j$ in $\LA_{KE}$ and the $f$ in $\LA_m$ are of the same order
of magnitude.\cite{ChiG} Only the value of the $f_3$ is fixed by
the relation
\be f_3 = \frac{v}{\sqrt{2}} \doteq 174\,{\rm GeV}
\label{f3}
\ee
following from the structure of the gauge boson mass matrix.

\section{Deviations from LET Due to PGB Exchanges\label{sec:pgbmod}}
\subsection{General discussion}
\noindent
\begin{table}[ht]
\tcaption{PGB exchange contributions to the individual processes in
different channels ($s,t,u$).
\label{gencontr}}
\begin{center}
\begin{tabular}{|c|c|c|c|c|c|}\hline
Process \# & Process & $s$ & $t$ & $u$  \\ \hline\hline
\rule[-3mm]{0cm}{8mm} 1 & $W^+_1 W^-_2 \to Z_3 Z_4$ & $\pi^0$ &
$\pi^+$ & $\pi^+$ \\ \hline
\rule[-3mm]{0cm}{8mm} 2 & $W^+_1 Z_2 \to Z_3 W^+_4$ & $\pi^+$ &
$\pi^+$ & $\pi^0$ \\ \hline
\rule[-3mm]{0cm}{8mm} 3 & $W^+_1 W^+_2 \to W^+_3 W^+_4$ &$\pi^{++}$&
$\pi^0$ & $\pi^0$  \\ \hline
\rule[-3mm]{0cm}{8mm} 4 & $W^+_1 W^-_2 \to W^+_3 W^-_4$ & $\pi^0$
& $\pi^0$ & $\pi^{++}$ \\ \hline
\end{tabular}
\end{center}
\end{table}
Table\,\ref{gencontr} gives an overview of general types
of PGB exchange contributions to the processes in question.
In the considered model there are six types of $\pi V V$
vertices, as one can see
from the relevant part of the interaction Lagrangian\cite{bahnik}
\ber
{\cal L}_{VV\pi} &=& -\alpha g m_W
\frac{1}{\sqrt{2}}\frac{f_{3'}}{f_5}\,W^+W^+\pi^5_{--} + c.c. \nn \\
&&+\,\alpha g m_W \sqrt{\frac{8}{3}}\frac{f_{3'}}{f_1}W^+W^- \,\pi^1
- \alpha g m_W\frac{1}{\sqrt{3}}\frac{f_{3'}}{f_5} W^+W^-\pi^5_0 \nn \\
&& -\,\frac{\alpha g m_Z}{\cos{\theta_W}}\frac{f_{3'}}{f_5} W^+Z \,\pi^5_- + c.c.
\nn \\
&&+\,\frac{\alpha g m_Z}{\cos{\theta_W}}
\frac{1}{\sqrt{3}}\frac{f_{3'}}{f_5}Z\,Z \pi^5_0 +
\frac{\alpha g m_Z}{\cos{\theta_W}}
\sqrt{\frac{2}{3}}\frac{f_{3'}}{f_1}Z\,Z \pi^1 \nn \\
&&+\,i g m_Z \sin^2{\theta_W}\,W^+Z\,G^- + c.c. - i e m_W\,W^+A\, G^- + c.c.
\label{lvvpi}\eer
where
\[\tan{\theta_W} = \frac{g'}{g},\quad m_W = \frac{g f_3}{\sqrt{2}},\quad m_Z =
\frac{f_3 \sqrt{g^2+g'^2}}{\sqrt{2}} \quad\mbox{and}\quad e=g\sin{\theta_W}
\ .\]
The fictitious GB $G^{\pm}$ are eliminated in the $U$-gauge. Unfortunately,
only the doubly charged field $\pi^5_{++}$ is a mass eigenstate
while the other $\pi$ fields, in general, are not. As an example,
we have performed numerical diagonalization of the GB mass
matrix for the values of parameters $\alpha = 0.5, 0.9$,
$a,b,c,d = 1$ and with all the $f'$s taken to be equal.\cite{bahnik}
The result has been
expressed in multiples of $f$. Using the only fixed value of $f$, $f_3$
in (\ref{f3}), we can get an estimate of the PGB masses. The mass
eigenstate fields $\varphi$ an fields $\pi$ are related by
\ber
\pi^1 &=& 0.107\, \varphi_7 + 0.997\,\varphi_8 \quad
    \pi^5_0 = -0.997\,\varphi_7 + 0.107\,\varphi_8 \nn \\
\pi^5_- &=& -i\,0.997\, \varphi_9 + i\,0.076\,\varphi_{11} \quad \pi^5_+ =
(\pi^5_-)^{\dagger}
\eer
The $\varphi_{7,8}$ and $\varphi_{9,11}$ are real and complex scalar
fields resp. In what follows we will neglect this fact and assume that the $\pi$
fields are themselves mass eigenstates.
To plot the relevant graphs we have chosen masses of $\pi^1$ and $\pi^5$
according to eigenvalues calculated in\cite{bahnik}
\be
m_{\pi^1} \doteq 3.1\,f \doteq 500\,{\rm GeV}
\quad
m_{\pi^5} \doteq 1.9\,f \doteq 350\,{\rm GeV}
\ee
with $f=f_3=174$\,GeV.

First we give general expressions of tree-level scattering amplitudes of
longitudinally polarized gauge bosons for each row of the
Table~\ref{gencontr}. Then we take into account the particle content of
the $SU(4)/SU(2)$ model. We will write the amplitude in a form
\be
 \M^{(n)}_{k\pi} = - g_1 g_2 A^{(n)}_{k\pi}, \quad k= s,t,u \ .
\ee
The couplings $g_1, g_2$ can be read off from (\ref{lvvpi}), e.g.
\be g_{WW\pi^5_{++}} = - 2\,\alpha g m_W \frac{1}{\sqrt{2}}\frac{f_{3'}}{f_5}
,\quad
g_{WW\pi^5_{0}} = - \alpha g m_W \frac{1}{\sqrt{3}}\frac{f_{3'}}{f_5},
\quad \ldots
\ee
Note the factor 2 in the coupling involving two identical particles. We
denote the contributions to the scattering amplitude comming from
PGB exchange as $\M^{(n)}_{SB}$. Then the complete amplitude of
the process \#$n$ is given by $\M^{(n)} = \M^{(n)}_{gauge} + \M^{(n)}_{SB}$. Exact
formulae for $\M^{(n)}_{gauge}$ are given in Ref.~23 and also in
the earlier papers Ref.~26.

\begin{figure}[t]
\begin{center}
\includegraphics[height=8cm,width=10cm]{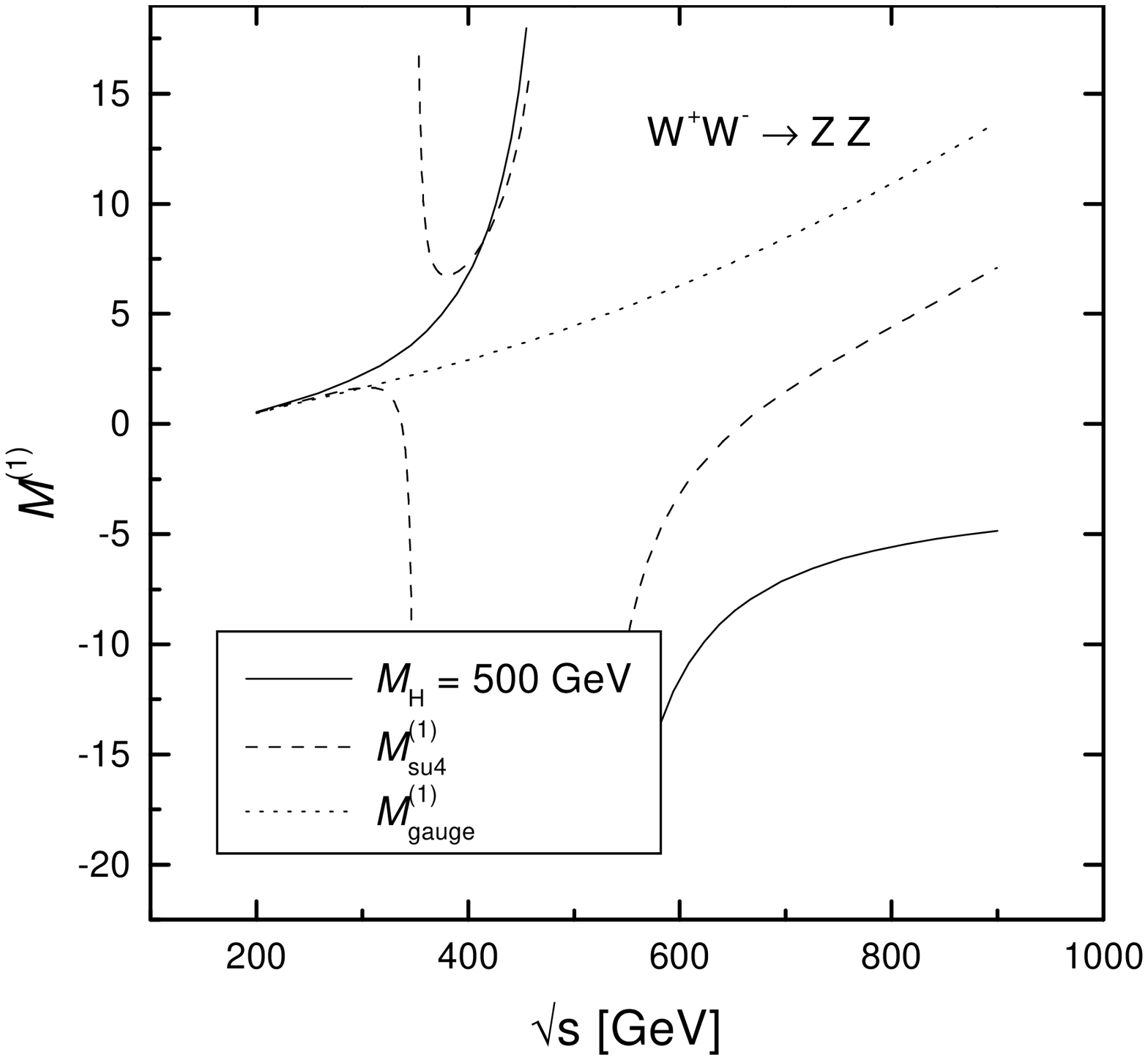}
\fcaption{Tree-level amplitude of the process $WW\to ZZ$
as a function of $\sqrt{s}$ in SM (solid), $SU(4)/SU(2)$ model (dashed),
and pure gauge amplitude $\M^{(1)}_{gauge}$ (dotted).
The Higgs boson mass $m_H = 500$\,GeV, width $\Gamma_H = 0$,
$\cos{\theta_{cm}} = 0.5$, $\rho=1$, $m_Z=91.2$\,GeV, $\sin^2{\theta_W} = 0.231$,
$g^2=\frac{4\pi\alpha_{em}}{\sin^2{\theta_W}}\doteq 0.42$,
$m_{\pi^5} = 350$\,GeV, $m_{\pi^1} = 500$\,GeV.\label{figwwzzsu}}
\end{center}
\end{figure}

\begin{figure}[t]
\begin{center}
\includegraphics[height=8cm,width=10cm]{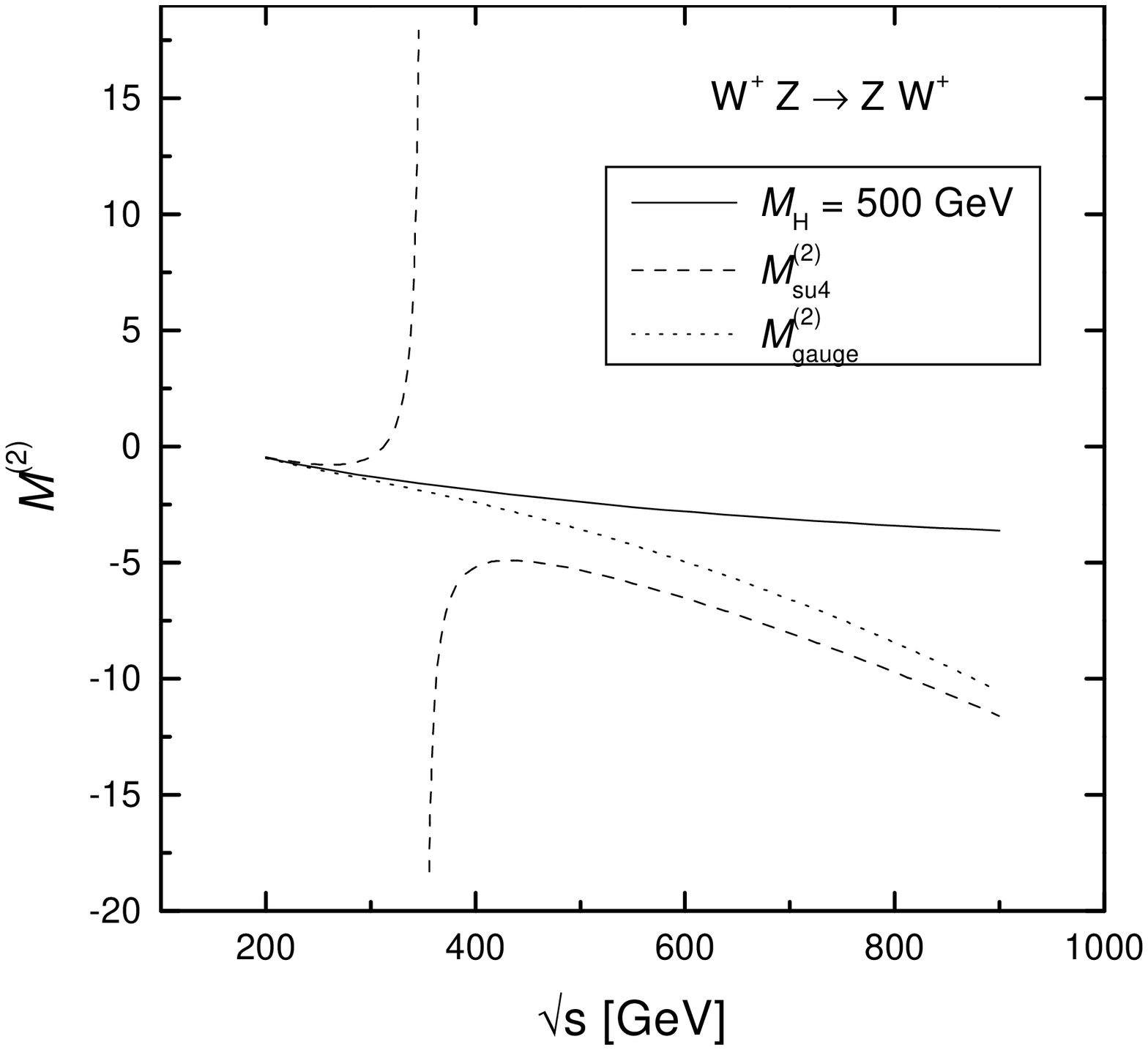}
\fcaption{Tree-level amplitude of the process $WZ\to ZW$
as a function of $\sqrt{s}$ in SM (solid), $SU(4)/SU(2)$ model (dashed),
and pure gauge amplitude $\M^{(2)}_{gauge}$ (dotted).
\label{figwzwzsu}}
\end{center}
\end{figure}

\begin{figure}[t]
\begin{center}
\includegraphics[height=8cm,width=10cm]{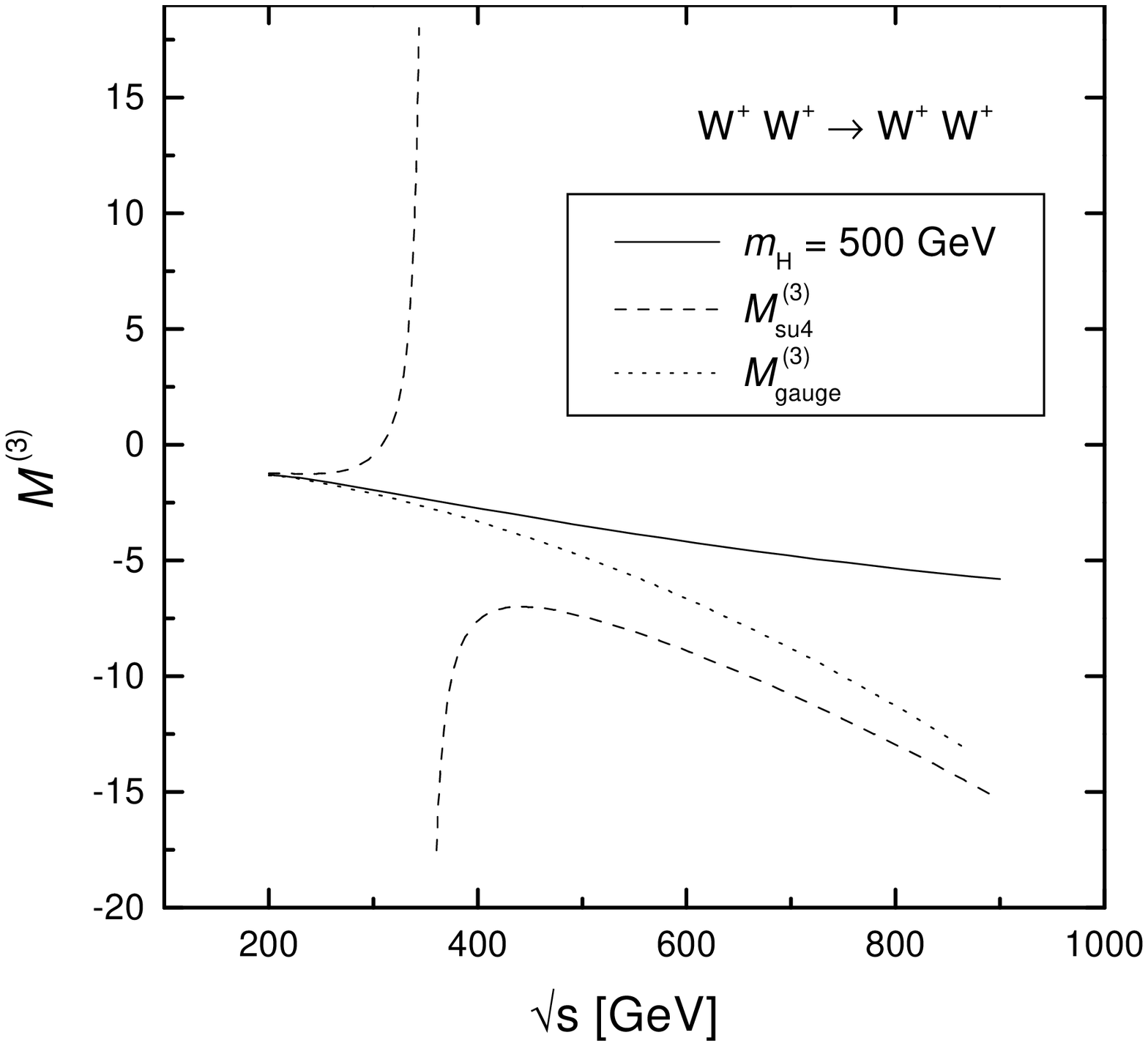}
\fcaption{Tree-level amplitude of the process $W^+W^+\to
W^+W^+$
as a function of $\sqrt{s}$ in SM (solid), $SU(4)/SU(2)$ model (dashed),
and pure gauge amplitude $\M^{(3)}_{gauge}$ (dotted).
\label{figwpwpsu}}
\end{center}
\end{figure}

\begin{figure}[t]
\begin{center}
\includegraphics[height=8cm,width=10cm]{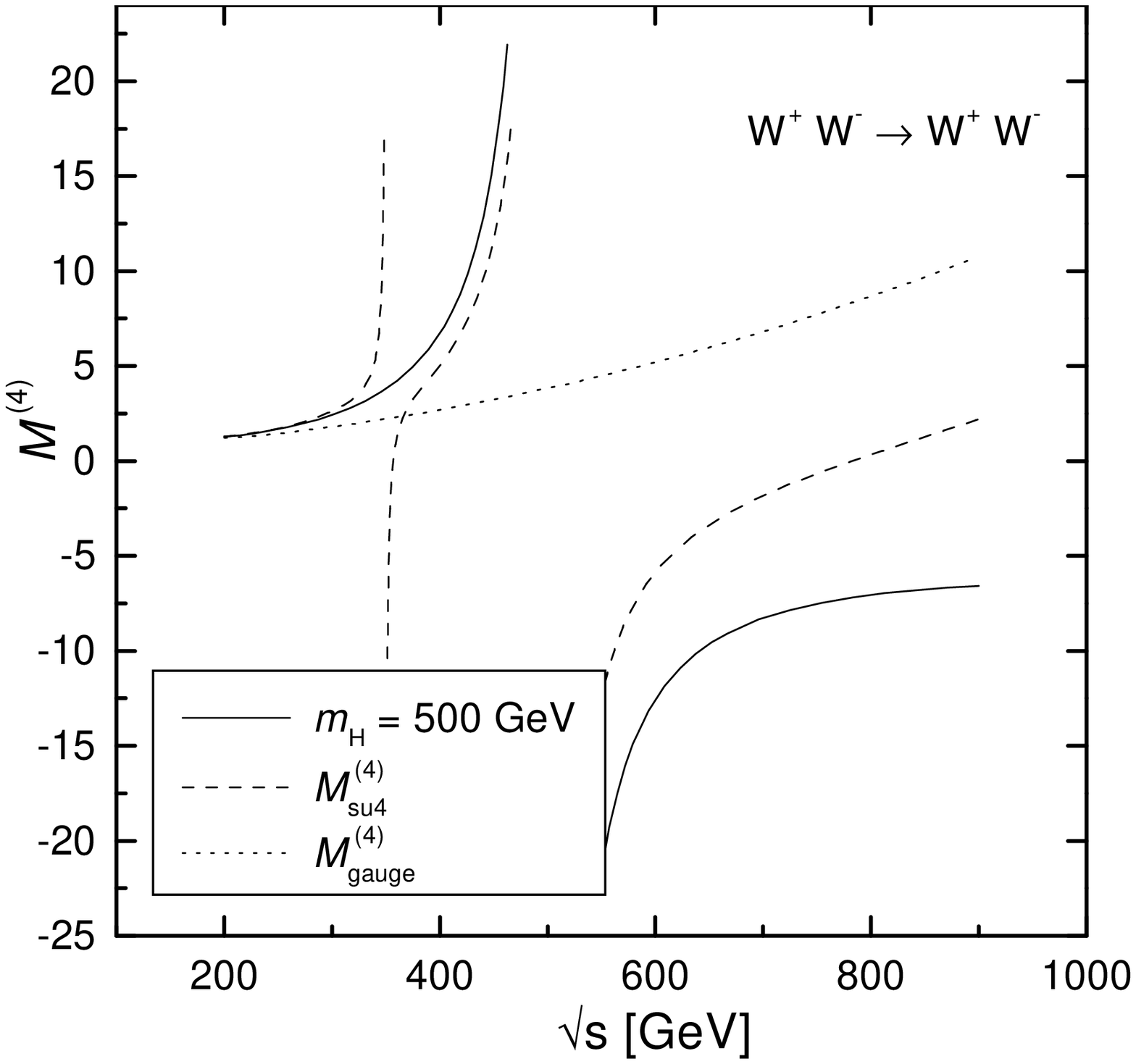}
\fcaption{Tree-level amplitude of the process $W^+W^-\to
W^+W^-$
as a function of $\sqrt{s}$ in SM (solid), $SU(4)/SU(2)$ model (dashed),
and pure gauge amplitude $\M^{(4)}_{gauge}$ (dotted).
\label{figwpwmsu}}
\end{center}
\end{figure}

Figures~\ref{figwwzzsu}-\ref{figwpwmsu} compare the tree-level
amplitudes corresponding to SM and the considered $SU(4)/SU(2)$
scheme. The Higgs boson mass is taken to be $m_H = 500$\,GeV.
The masses of PGB $\pi^5$ and $\pi^1$ are $m_{\pi^5} = 350$\,GeV
and $m_{\pi^1} = 500$\,GeV resp. as discussed above, $\alpha=0.5$ and
$f_{3'} = f_5 = f_1$. The parameters are the same for all the plots. Pure
gauge amplitudes $\M^{(n)}_{gauge}$ (dotted line) are also plotted.  As
mentioned in Section~2, the high-energy limit of the $\M^{(n)}_{gauge}$
practically coincides with the amplitude given by the canonical LET.

\subsection{Results for Scattering Amplitudes}
\noindent
Using values of coupling constant from (\ref{lvvpi}) and high-energy
limits of $A'$s given in (\ref{heexp}) we get for the process \#1
\be
\M^{(1)}_{SB} =
-\frac{\alpha^2\,g^2}{m_W^2}\left[\left(\frac{f_{3'}}{f_1} \right)^2
\frac{2 s \sqrt{\rho}}{3} + \left(\frac{f_{3'}}{f_5} \right)^2
\left(\frac{m_Z^2 \rho}{4 m_W^2}(t+u) - \frac{s \sqrt{\rho}}{6}\right)\right]
+ O(s^0)
\label{m1sb}
\ee
where the contributions from individual terms (see (\ref{eq:m1sbterms}))
are clearly distinguished. Of course,
in the considered model $\rho = m_W^2 / m_Z^2
\cos^{2}{\theta_W} = 1$. For the simplest case $f_{3'} = f_5 = f_{1}$
and using $s + t + u \doteq 0$ we get
\[ \M^{(1)}_{SB} = -\frac{\alpha^2 g^2 s}{m_W^2}\left[\frac{1}{2} -
\frac{m_Z^2}{4 m_W^2}\right] + O(s^0)
\]
In the case of the process \#2 the symmetry-breaking amplitude differs form
$\M^{(1)}_{SB}$ by replacement $s\leftrightarrow u$. Taking into account
only the leading terms in $s$ we get an analogous expression as for
$\M^{(1)}_{SB}$
\be
\M^{(2)}_{SB} =
-\frac{\alpha^2\,g^2}{m_W^2}\left[\left(\frac{f_{3'}}{f_1} \right)^2
\frac{2 u \sqrt{\rho}}{3} + \left(\frac{f_{3'}}{f_5} \right)^2
 \left(\frac{m_Z^2 \rho}{4 m_W^2}(s+t) - \frac{u \sqrt{\rho}}{6}\right)\right]
 + O(s^0) \ .
 \label{m2sb}
\ee
Similarly, for our simple choice of parameters $f$
\[ \M^{(2)}_{SB} = -\frac{\alpha^2 g^2 u}{m_W^2}\left[\frac{1}{2} -
\frac{m_Z^2}{4 m_W^2}\right] + O(s^0) \ .
\]
\noindent
The relevant formulae for processes \#3 and \#4 are easily obtained by
setting $m_Z=m_W$ in (\ref{a1tpi}) or (\ref{a2tpi}).
\be \M^{(3)}_{SB} =
-\frac{\alpha^2\,g^2}{m_W^2}\left[\frac{1}{12}\left(\frac{f_{3'}}{f_5} \right)^2
(6 s + t + u) + \frac{2}{3}\left(\frac{f_{3'}}{f_1} \right)^2
 (t + u) \right] + O(s^0)
\label{m3sb}
\ee
which for our choice of the $f$'s simplifies to
\[\M^{(3)}_{SB} = \frac{\alpha^2\,g^2 s}{4m_W^2} + O(s^0) \ .\]
\be \M^{(4)}_{SB} =
-\frac{\alpha^2\,g^2}{m_W^2}\left[\frac{1}{12}\left(\frac{f_{3'}}{f_5} \right)^2
(6 u + t + s) + \frac{2}{3}\left(\frac{f_{3'}}{f_1} \right)^2
 (s + t) \right] + O(s^0)
\label{m4sb}
\ee
and for $f_{3'} = f_5 = f_{1}$ this becomes
\[\M^{(4)}_{SB} = \frac{\alpha^2\,g^2 u}{4m_W^2} + O(s^0)\]

\section{Discussion of Results}
\noindent
Using the formulae (\ref{m1sb}), (\ref{m2sb}), (\ref{m3sb}) and (\ref{m4sb}) we can
discuss the modifications of the low-energy theorems
caused by PGB exchanges. Table\,\ref{tab:pgbcom}
displays the high-energy limit ($s\gg
m_W^2,m_{\pi}^2, m_H^2$) of the symmetry
breaking amplitudes in the extended model (up to the factor
$-\frac{\alpha^2 g^2 f_{3'}^2}{4m_W^2}$) and in the SM with a light Higgs
boson (up to the factor $-\frac{g^2}{4 m_W^2}$). The approximate relation
$s+t+u\doteq 0$ is used.

\begin{table}[ht]
\tcaption{A comparison of high-energy limit of the individual
PGB and Higgs
contributions to the scattering amplitudes of $W_L$ and $Z_L$.
$\rho =1$, $s +t+u \doteq 0$.\label{tab:pgbcom}}
\begin{center}
\begin{tabular}{|c|c|c|c|c|c|}\hline
\rule[-3mm]{0cm}{8mm} Process & $\pi^5_{++}$ & $\pi^5_{+}$ & $\pi^5_{0}$ &
$\pi^1$ & SM Higgs \\ \hline\hline
\rule[-3mm]{0cm}{8mm} $W^+_1 W^-_2 \to Z_3 Z_4$ & &
$-\frac{s}{\cos^2{\theta_W} f_5^2}$ & $-\frac{s}{f_5^2}$ & $\frac{8s}{3 f_1^2}$&
$s$ \\ \hline
\rule[-3mm]{0cm}{8mm} $W^+_1 Z_2 \to Z_3 W^+_4$ & &
$-\frac{u}{\cos^2{\theta_W} f_5^2}$ & $-\frac{u}{f_5^2}$ & $\frac{8u}{3 f_1^2}$&
$u$ \\ \hline
\rule[-3mm]{0cm}{8mm} $W^+_1 W^+_2 \to W^+_3 W^+_4$ &
$\frac{2s}{f_5^2}$& & $-\frac{s}{3 f_5^2}$ & $-\frac{8s}{3 f_1^2}$&
$-s$ \\ \hline
\rule[-3mm]{0cm}{8mm} $W^+_1 W^-_2 \to W^+_3 W^-_4$ &
$\frac{2u}{f_5^2}$& & $-\frac{u}{3 f_5^2}$ & $-\frac{8u}{3 f_1^2}$&
$-u$ \\ \hline
\end{tabular}
\end{center}
\end{table}
As regards the process $ZZ\to ZZ$, neutral PGB exchanges
in $s$, $t$ and $u$ channels contribute there. In the limit $s,t,u \gg m_W, m_Z,
m_{\pi}$ one thus gets an amplitude proportional to $s+t+u
\sim 0$ (see (\ref{heexp})). The signs of neutral PGB contributions to the processes \#3 and
\#4 are the same as the Higgs' ones. It is interesting that no
modifications occur for $\alpha=0$. The value of $\alpha$ is
restricted\cite{ChiG} by the GB kinetic term to $|\alpha|<1$. The presence
of a term proportional to $\alpha$ is due to the existence of two {\em
different} triplets $3$ and $3'$ (see (\ref{lke})). The fact that there
can be more than one $H$-irreducible multiplet was pointed out
in Refs.~27,~28, where a most general Lagrangian is written in
the form
\be
\LA_{BB} = \sum^{n}_{i=1} f_i^2 J^i_{\mu} \cdot J^{i\mu}
\ee
with $n$ being the number of $H$-irreducible multiplets. Thus the fact
that it is possible to combine {\em different} multiplets was neglected.
A detailed discussion of the structure of an effective chiral
Lagrangian containing
combinations of different multiplets along with the classification of
GB and the corresponding definition of the $U$-gauge can be found in
Ref.~23.
Table\,\ref{tab:compact} gives a more compact view on the PGB contribution in the case
$f_{3'} = f_5 = f_{1}$.
\begin{table}[ht]
\tcaption{A comparison of the LET,
the high-energy limit of Higgs-boson and the PGB contributions
with $f_{3'} = f_5 = f_{1}$.
\label{tab:compact}}
\begin{center}
\begin{tabular}{|c|c|c|c|}\hline
\rule[-3mm]{0cm}{8mm} Process & LET & SM Higgs & $SU(4)/SU(2)$\\ \hline\hline
\rule[-3mm]{0cm}{8mm} $W^+_1 W^-_2 \to Z_3 Z_4$ & $\frac{g^2 s}{4 \rho m_W^2}$
&$-\frac{g^2 s\sqrt{\rho}}{4 m_W^2}$ &
$-\frac{\alpha^2 g^2 s\sqrt{\rho}}{m_W^2}\left[\frac{1}{2} -
\frac{m_Z^2\sqrt{\rho}}{4 m_W^2}\right]$  \\ \hline
\rule[-3mm]{0cm}{8mm} $W^+_1 Z_2 \to Z_3 W^+_4$ &$\frac{g^2 u}{4 \rho m_W^2}$
& $-\frac{g^2 u\sqrt{\rho}}{4 m_W^2}$ &
$-\frac{\alpha^2 g^2 u\sqrt{\rho}}{m_W^2}\left[\frac{1}{2} -
\frac{m_Z^2\sqrt{\rho}}{4 m_W^2}\right] $  \\ \hline
\rule[-3mm]{0cm}{8mm} $W^+_1 W^+_2 \to W^+_3 W^+_4$
& $-\frac{g^2 s}{4 m_W^2}\left(4 - \frac{3}{\rho}\right )$
&$\frac{g^2 s}{4 m_W^2}$  &
$\frac{\alpha^2\,g^2 s}{4m_W^2}$  \\ \hline
\rule[-3mm]{0cm}{8mm} $W^+_1 W^-_2 \to W^+_3 W^-_4$
&$-\frac{g^2 u}{4 m_W^2}\left(4 - \frac{3}{\rho}\right)$ &$\frac{g^2 u}{4 m_W^2}$  &
$\frac{\alpha^2\,g^2 u}{4m_W^2} $ \\ \hline
\end{tabular}
\end{center}
\end{table}
Comparing Fig.\,\ref{figwwzzsu} ($WW\to ZZ$) and
Fig.\,\ref{figwpwmsu} ($W^+W^-\to W^+W^-$) we can see a
manifestation of the opposite signs of the couplings
\be
g_{ZZ\pi^5_0} = 2\,\frac{\alpha g m_Z}{\cos{\theta_W}}
\frac{1}{\sqrt{3}}\frac{f_{3'}}{f_5} \quad\mbox{and}\quad
g_{WW\pi^5_0} =  - \frac{\alpha g m_Z}{\sqrt{3}}\frac{f_{3'}}{f_5}
\ee
(see (\ref{lvvpi})) around the pole at $\sqrt{s} =
m_{\pi^5} = 350$\,GeV. In the region $s>m_{\pi^1}, m_H = 500$\,GeV the
effect of the PGB exchange points in the same direction as that
of the SM Higgs boson. However, the
growth of the pure gauge amplitudes is not damped by the PGB
exchange, at least for our simple choice of the $f'$s.

Looking at Fig.~\ref{figwzwzsu} ($WZ\to ZW$) and Fig.~\ref{figwpwpsu}
($W^+W^+\to W^+W^+$) one may notice that the PGB act in an opposite
way than the
Higgs boson, in contradiction with the last row of the
Table\,\ref{tab:compact}. But the energies covered by the figures are
not in the region $\sqrt{s}\gg m_{\pi}, m_H$. There is a point near
1200\,GeV where $\M^{(2)}_{SB}$ and $\M^{(3)}_{SB}$ change the sign.
Nevertheless they cannot cancel the ``bad'' high-energy behaviour of the
pure gauge amplitude.

\section{Concluding Remarks}
\noindent
The analysis performed in the present paper serves mainly an
illustrative purpose. We have shown, by means of an explicit
calculation, that a ``canonical'' low-energy
theorem\cite{ChanowitzLET} for strongly
interacting $W_{L}$ and $Z_{L}$, proven for a ``minimal''
symmetry-breaking scenario, may indeed be violated
within an extended scheme involving physical PGB with masses on
the electroweak scale.

The considered scheme (introduced earlier\cite{ChiG} in a slightly
different context), corresponding to spontaneous breaking of the
symmetry $SU(4)$ down to (custodial) $SU(2)$ subgroup, is somewhat
unusual in that the Goldstone-boson manifold $G/H = SU(4)/SU(2)$ is
not a symmetric space. For reader's convenience let us recall
that a $G/H$ (the quotient space for a group $G$ broken down to
$H$) is
symmetric space if there is a ``parity'' operation (an involutive
automorphism) $P$ on the Lie algebra of the $G$, distinguishing the
unbroken and broken generators $t_{i}$ and $x_{a}$ in such a way that e.g.
\begin{equation}
\label{parita}
P t_{i} P = + t_{i}, \qquad P x_{a} P = - x_{a}
\end{equation}
A useful criterion (a necessary condition) for the existence of a
$P$ satisfying (\ref{parita}) is that any commutator of the broken
generators can be expressed as a linear combination of the
unbroken ones.\cite{Peskin} For the SB scheme discussed in the
present paper one can show that this criterion is indeed violated
(in particular, a commutator of the generators belonging to the
two different triplets does not have the required property). In
this connection it is quite instructive to realize that several
rather broad classes of EWSB schemes involving rich spectra of PGB have
been studied previously\cite{Peskin,Chadha,Preskill}, mostly
within the framework of technicolour models. As noticed in the
Introduction, in any of these models the corresponding Goldstone-boson
manifold $G/H$ is a symmetric space and the PGB interactions
do not influence the canonical low-energy theorem.

One should also notice that the scheme examined here has
similar signatures as some weak-coupling models with
additional Higgs triplets investigated by other
authors\cite{Machacek,Vega},
in that doubly charged scalar bosons appear in the
physical spectrum. However, at present it is not clear to us how to
construct e.g. a Higgs potential that would reproduce precisely
the SB pattern considered in this paper. A discussion of such a
problem, as well as a comparison of scattering amplitudes
calculated here with those obtained e.g. in the model of Ref.~22
would deserve a separate treatment.

\hspace{1cm}

\noindent
{\bf Acknowledgements}

\noindent
This work has been partially supported by the grant GACR--202/98/0506.
One of us (J.H.) is indebted to Dr. M.St\"{o}hr for technical
assistance.

\nonumsection{References}

\begin{appendix}

\noindent
This appendix summarizes intermediate formulae used in our
calculations.\\[3ex]
{\bf Process $W^+(k_1) +  W^-(k_2) \to Z(k_3)+ Z(k_4)$}
\[A^{(1)}_{s\pi}(s) = \frac{(\ep_1\cdot\ep_2) (\ep_3^*\cdot\ep_4^*)_{long}}
{s - m_{\pi}^2} = \frac{(s - 2m_W^2)(s - 2m_Z^2)}{4m_W^2m_Z^2(s -
m_{\pi}^2)}
\]
\be A^{(1)}_{t\pi}(s,\cos{\theta_{cm}}) = \frac{(\ep_1\cdot\ep_3^*) (\ep_2\cdot\ep_4^*)_{long}}{t -
m_{\pi}^2} = \frac{(s\beta_W^2 - 4m_W^2\beta_Z^2 - s
\beta_W\beta_Z\cos{\theta_{cm}})^2}{16m_W^2 m_Z^2\beta_W^2\beta_Z^2 (t -
m_{\pi}^2)}
\label{a1tpi}
\ee
\[A^{(1)}_{u\pi}(s,\cos{\theta_{cm}}) = \frac{(\ep_1\cdot\ep_4^*)
(\ep_2\cdot\ep_3^*)_{long}}{u - m_{\pi}^2}= A^{(1)}_{t\pi}(s,-
\cos{\theta_{cm}})\]
where
\ber t &=& m_W^2 + m_Z^2 - \frac{s}{2} + \frac{s}{2}\,\beta_W\beta_Z \cos{\theta_{cm}}\nn\\
     u &=& m_W^2 + m_Z^2 - \frac{s}{2} - \frac{s}{2}\,\beta_W\beta_Z
     \cos{\theta_{cm}}\nn
\eer
\[\beta_W = \sqrt{1 - \frac{4 m_W^2}{s}}\; , \qquad \beta_Z = \sqrt{1 -
    \frac{4 m_Z^2}{s}}
 \ .\]
$\theta_{cm}$ is the angle between $\bk_1$ and $\bk_3$ in the
c.m. system.
Since we assume that the PGB are light, i.e. $m_{\pi}\sim m_W$, the high-energy ($s\gg m_W^2$)
expansion of the amplitudes takes a
simple form
\be A^{(1)}_{k\pi} = \frac{k}{4m_W^2m_Z^2} + O(s^0), \qquad
 k=s,t,u
\label{heexp}
\ee
\[ t = -\frac{s}{2}(1 - \cos{\theta_{cm}}),\
  u = -\frac{s}{2}(1 + \cos{\theta_{cm}}) \]
The symmetry-breaking part of the complete amplitude is
\be \M^{(1)}_{SB} =   \M^{(1)}_{s\pi^1} + \M^{(1)}_{s\pi^5_0} +
\M^{(1)}_{t\pi^5_+} +  \M^{(1)}_{u\pi^5_+}\ .
\label{eq:m1sbterms}
\ee
{\bf Process $W^+(k_1)+ Z(k_2) \to Z(k_3) + W^+(k_4)$}
\[A^{(2)}_{s\pi}(s) = \frac{(m_W^2 + m_Z^2 - s)^2}{4m_W^2m_Z^2 (s -
m_{\pi}^2)} \]
\be A^{(2)}_{t\pi} = \frac{[3(m_W^2 - m_Z^2)(m_W^4 - m_Z^4) - (m_W^2 - m_Z^2)^2 t
+ s^2 m_W^2 \beta_W^2 + s^2 m_Z^2 \beta_Z^2 + s^2 t]^2}{4m_W^2 m_Z^2 [(m_W -m_Z)^2
-s]^2 [(m_W+m_Z)^2 - s]^2(t-m_{\pi}^2)}
\label{a2tpi}
\ee
\[A^{(2)}_{u\pi}=\frac{C^{(2)}_u}{J^{(2)}_u} \]
\begin{eqnarray*}
C^{(2)}_u = \left[(m_W^2 -m_Z^2)^2 (2m_W^2 + u) + 2 (m_W^2 -
m_Z^2) s u\right.\\
\left. {}+ 2 m_W^2 s (s - 2 (m_W^2 + m_Z^2)) + s^2u\right] [m_W\leftrightarrow m_Z]
\end{eqnarray*}
\[J^{(2)}_u=4 m_W^2 m_Z^2 [(m_W - m_Z)^2 - s]^2 [(m_W + m_Z)^2 - s]^2 ( u-
m_{\pi}^2)
\]
The kinematical variables are related by
\[t = m_W^2 + m_Z^2 -\frac{s}{2} + \frac{(m_Z^2 - m_W^2)^2}{2 s} + 2\,k^2
\cos{\theta_{cm}} \]
\[u = - 2 k^2 (1 + \cos{\theta_{cm}}) \]
where in the c.m. system
\[ k^2 = \frac{1}{4 s}\left [s^2 + (m_W^2 - m_Z^2)^2 - 2s
\,(m_W^2 +m_Z^2)\right ] = |\bk_i|^2 \quad i = 1,2,3,4. \]
High-energy expansions are the same as in (\ref{heexp}).\\[3ex]
{\bf Process $W^+(k_1)+ W^+(k_2) \to W^+(k_3) + W^+(k_4)$}\\
The relevant formulae for processes \#3 and \#4 are easily obtained by
setting $m_Z=m_W$ in (\ref{a1tpi}) or (\ref{a2tpi}).
\[A^{(3,4)}_{s\pi} = \frac{(2m_W^2 - s)^2}{4m_W^4 (s - m_{\pi}^2)} \]
\[A^{(3,4)}_{k\pi} = \frac{(2m_W^2\beta_W^2 + k)^2}{4m_W^4\beta_W^4(k -
m_{\pi}^2)} \qquad k = t,u
\]
\be
\M^{(3)}_{SB} =   \M^{(3)}_{s\pi^5_{++}} + \M^{(3)}_{t\pi^5_0} +
\M^{(3)}_{u\pi^5_0} +  \M^{(3)}_{t\pi^1}  + \M^{(3)}_{u\pi^1}\ .
\ee
{\bf Process $W^+(k_1)+ W^-(k_2) \to W^+(k_3) + W^-(k_4)$}\\
For the sake of completeness we give also the corresponding results
for the process \#4
\be
\M^{(4)}_{SB} =   \M^{(4)}_{u\pi^5_{++}} + \M^{(4)}_{t\pi^5_0} +
\M^{(4)}_{s\pi^5_0} +  \M^{(4)}_{t\pi^1}  + \M^{(4)}_{s\pi^1}\ .
\ee
\end{appendix}

\end{document}